\documentclass[12pt,subeqn,a4paper]{article}

\usepackage{amssymb,amsmath,amsfonts,amsthm, amscd, mathrsfs,helvet}
\usepackage{bm}
\usepackage{graphicx,verbatim}
\usepackage{psfrag}
\usepackage[all]{xy}

\def\r2{\sqrt{2}}

\def\bea{\begin{eqnarray} }
\def\eea{\end{eqnarray}}
\def\be{\begin{equation} }
\def\ee{\end{equation}}
\def\nn{\nonumber}

\begin{document}
\newcommand{\nd}[1]{/\hspace{-0.5em} #1}
\begin{titlepage}
\begin{flushright}
{\bf January 2010} \\ 
\end{flushright}
\begin{centering}
\vspace{.2in}
 {\large {\bf Giant Holes}}

\vspace{.3in}

Nick Dorey and Manuel Losi\\
\vspace{.1 in}
DAMTP, Centre for Mathematical Sciences \\ 
University of Cambridge, Wilberforce Road \\ 
Cambridge CB3 0WA, UK \\
\vspace{.2in}
%
%
\vspace{.4in}
{\bf Abstract} \\
\end{centering}

We study the semiclassical spectrum of excitations 
around a long spinning string in $AdS_{3}$. 
In addition to the usual small fluctuations, we find the
spectrum contains a branch of solitonic excitations of finite energy. 
We determine the dispersion relation for these excitations. This 
has a relativistic form at low energies but also matches 
the dispersion relation for the ``holes'' of the dual gauge theory spin
chain at high energies. The low-energy behaviour is consistent with
the hypothesis that the solitonic excitations studied here 
are continuously related to the elementary excitations of the string.
   

\end{titlepage}
\paragraph{}
\paragraph{}

The discovery of the integrability of the
one-loop planar dilatation operator in 
${\cal N}=4$ SUSY Yang-Mills \cite{MZ, B} and of the classical limit
of the dual string theory on $AdS_{5}\times S^{5}$ \cite{BPR}
initiated a period of rapid 
progress in our understanding of the AdS/CFT correspondence. Recently 
this progress has culminated in 
a set of equations which are conjectured to describe the exact spectrum of
both theories in the planar limit \cite{GKV, BFT, AF1}, for all values of
the 't Hooft coupling $\lambda=g^{2}_{YM}N$. So far most of this progress is
based on our knowledge of the exact spectrum of excitations around the 
the one-half BPS chiral primary operator\footnote{Here $Z$ is a 
complex adjoint scalar of the ${\cal N}=4$ theory.} 
$\hat{O}_{\rm BPS}={\rm Tr}[Z^{J}]$ in
the $J\rightarrow \infty$ large-volume limit 
and of the exact two body S-matrix for the
scattering of these excitations \cite{Staudacher, BS, B1, BES}. 
In particular, knowledge
of this large-volume data is the starting point for the Y-system and 
Thermodynamic
Bethe Ansatz approaches to the finite volume theory.  
\paragraph{}
The BPS chiral primary operator described above corresponds to a
point-like string orbiting the equator of the five-sphere in the dual
geometry; also known as the BMN groundstate \cite{BMN}. 
As recently emphasized in \cite{AM2, AMs}, an alternative starting point 
or ``ground-state'' for quantization of the string is provided by the 
folded spinning string 
solution of Gubser, Klebanov and Polyakov \cite{GKP}. This is the
state of lowest dimension\footnote{In the following we will adopt the
  gauge theory terminology and refer to $\Delta$, the Noether charge
  corresponding to translation of global time in $AdS$, as dimension. 
The term energy will be reserved for the combination 
$E=\Delta-S$.} $\Delta$ with a single fixed spin, $S$, in 
$AdS_{5}$. In the $S\rightarrow \infty$ limit, the energy of this 
configuration is proportional to the 
cusp anomaly $\Gamma(\lambda)$ of the ${\cal N}=4$
theory and it is also a natural starting point for the connection to
light-like Wilson loops and gluon scattering amplitudes discussed by
Alday and Maldacena \cite{AMs, KrTs}. 
As in the $J\rightarrow\infty$ limit of the BMN groundstate, 
the string effectively becomes infinitely long in 
this large spin limit. As the worldsheet theory is integrable 
we expect that excitations of finite energy around 
this ground-state will correspond to particles which undergo factorised
scattering. The main purpose of this note is to study these
excitations in semiclassical string theory and compare them with the
corresponding states in the dual gauge theory.  
\paragraph{}
In the following we will focus on excitations with energy of
order $\sqrt{\lambda}$ which appear as classical solutions of the
worldsheet theory. A large class of solutions of this type has been
constructed in \cite{Jev1} using the inverse scattering method. 
We will show that a particular subset of these solutions correspond to
local excitations of the GKP spinning string which make a finite
contribution to the energy $E=\Delta-S$ and we 
will investigate the dynamics of these states. 
Our main result is an explicit expresion for the
dispersion relation of these excitations. We find that the energy and
momentum of the soliton are related as,    
\bea E_{\rm sol}(v) & = & \frac{\sqrt{\lambda}}{2\pi}
\left[\frac{1}{2}\log\left(\frac{1+\sqrt{1-v^{2}}}
{1-\sqrt{1-v^{2}}}\right)\,\,-\,\,
\sqrt{1-v^{2}}\right] \nn \\ 
P_{\rm sol}(v) & = &
\frac{\sqrt{\lambda}}{2\pi}\left[\frac{\sqrt{1-v^{2}}}{v}\,\,-\,\,
{\rm Tan}^{-1}\left(\frac{\sqrt{1-v^{2}}}{v}\right)\right] 
\label{disp2} \eea
where the branch of inverse tangent is chosen so that $-\pi/2\leq {\rm
  Tan}^{-1} x\leq +\pi/2$ for all real values of $x$. 
Here $E_{\rm sol}$ is the contribution of the soliton to the energy 
$E=\Delta-S$. 
The momentum $P_{\rm sol}$ is canonically conjugate to the position
of the soliton on the string in coordinates where the total length of the
string grows like $2\log S$ as $S\rightarrow \infty$. The parameter 
$v\leq 1$ is
the velocity of the soliton in these coordinates. 
\paragraph{}
The physical information in the above dispersion relation is the semiclassical
spectrum of states associated with the excitation. As we describe
below, this corresponds to the relation $E_{\rm sol}(P_{\rm sol})$,
as determined by (\ref{disp2}), supplemented with 
the large-$S$ quantization condition, 
\bea
P_{\rm sol}\cdot\, 2\log S & \in & 2\pi\,\,\mathbb{Z} \nn \eea
The resulting spectrum can also be obtained 
from a semiclassical quantization of the finite gap
construction of \cite{KMMZ}. 
A more complete analysis will appear in the forthcoming
paper \cite{DL}. 
\paragraph{}   
At low momentum ($v\simeq 1$), the dispersion relation takes the
relativistic form, 
\bea 
E_{\rm sol} & \simeq & |P_{\rm sol}|\,\, +\,\, {\rm
  O}\left(P^{\frac{5}{3}}_{\rm sol}\right) 
\nn \eea 
This is similar to the behaviour of the Giant Magnons \cite{HM} 
which appear as
solitonic excitations of the BMN groundstate. In that case the onset
of relativistic behaviour at low momenta allows the identification of
Giant Magnons with the elementary excitations of the string. In the
following we will argue that a similar phenomenon should also occur in this
case. Away from low energy the dispersion relation is
non-relativistic. Although the conformal gauge string action and
Virasoro constraint respect worldsheet Lorentz invariance, this
symmetry is broken by fixing the residual gauge symmetry and the
non-relativistic dispersion relation of the soliton reflects 
this breaking.     
\paragraph{}
The spinning folded string in $AdS_{3}$ is dual to a certain twist two
operator in ${\cal N}=4$ SUSY Yang-Mills. 
Excitations around this operator correspond to
holes in the rapidity distribution in the corresponding integrable
spin chain \cite{BGK,FS}. These holes have the characteristics of
particles carrying conserved momenta 
and each such hole contributes a finite amount to the anomalous
dimension of the operator. Taking our dispersion relation
(\ref{disp2}) in the limit of large momentum ($v<<1$) we find, 
\bea 
E_{\rm sol} & = & \frac{\sqrt{\lambda}}{2\pi}\,\log\,|P_{\rm
  sol}|\,\, + \,\,  {\rm
  O}\left(P^{0}_{\rm sol}\right)    
\label{largep} \eea
As we review below, this precisely matches the relevant dispersion
relation for so-called ``large'' holes up to the by now familiar 
replacement of 
the prefactor ${\sqrt{\lambda}}/{2\pi}$ with the cusp anomalous dimension 
$\Gamma(\lambda)$. This is one of several examples where  
gauge theory and string theory quantities agree modulo this replacement
in the limit of large spin. In view of this agreement we suggest that    
the string solitons studied in this paper deserve the name 
``Giant Holes''. The implied relation between holes and classical
solitons is reminiscent of similar phenomena occuring in the case of the
non-linear Schrodinger equation \cite{NLS}\footnote{We thank 
K. Zarembo for drawing this reference to our attention.}. 
 
\paragraph{}
We will now review the contruction of explicit solutions of the string
equations of motion in $AdS_{3}$ \cite{DeVega, Jev1}. 
We start by representing the target
space as a 3-dimensional hyperboloid 
embedded in $\mathbb{R}^{2,2}$ defined by the following constraint:
\begin{equation}
X_\mu X^\mu = -X_0^2-X_1^2+X_2^2+X_3^2 = -1
\label{eq:AdS3constr}
\end{equation}
The $\sigma$-model action, in conformal gauge, is then defined 
in terms of the embedding coordinates $X_\mu$ as:
\begin{equation}
I = - \frac{\sqrt{\lambda}}{4\pi}\int d\sigma d\tau \left[ 
G_{\mu\nu} \partial_a X^\mu \partial^a X^\nu + \Lambda \left( 
X_\mu X^\mu + 1 \right) \right]
\label{eq:AdS3sigmamodelaction}
\end{equation}
where $\lambda$ is the t'Hooft coupling and 
$G_{\mu\nu} = \mathrm{diag}(-1,-1,1,1)$ is the flat metric on 
$\mathbb{R}^{2,2}$. As we consider string motion in $AdS_{3}$ only,
the Virasoro constraint takes the form, 
\begin{equation}
\partial_\pm X^\mu \partial_\pm X_\mu = 0
\label{eq:Virasoropm}
\end{equation}  
\paragraph{}
The embedding coordinates defined in (\ref{eq:AdS3constr}) 
are related to the standard global
coordinates in $AdS_{3}$ via the complex combinations,
\begin{eqnarray}
 Z_1 & = & X_0 + i X_1 = \cosh\rho \; e^{i t} \nonumber \\
 Z_2 & = & X_2 + i X_3 = \sinh\rho \; e^{i \phi}
\label{eq:AdS3defcxcoords}
\end{eqnarray}
where $t$ is global time and $\phi\sim \phi + 2\pi$ is an angular
coordinate. 
As mentioned above, 
we will refer to the Noether charge conjugate to
translations of the global time $t$ as dimension $\Delta$.
Angular momentum $S$ in $AdS_{3}$ is canonically 
conjugate to translations of $\phi$. 
In terms of the string coordinates we have, 
\begin{eqnarray}
 \Delta & = & \frac{\sqrt{\lambda}}{2\pi} \int d\sigma \; \mathrm{Im}(\bar{Z}_1 \partial_\tau Z_1) \label{eq:AdS3_energy} \\
 S & = & \frac{\sqrt{\lambda}}{2\pi} \int d\sigma \; 
\mathrm{Im}(\bar{Z}_2 \partial_\tau Z_2) \label{eq:AdS3_spin} \eea
where the integrals are taken over the appropriate range of the
worldsheet coordinate $\sigma$. 

\paragraph{}
Pohlmeyer's reduction procedure \cite{Pohlmeyer, other} 
starts by defining a set of basis
vectors $\{B_{\mu}, X_{\mu},$ $\partial_+ X_{\mu},\partial_- X_{\mu}\}$
on $\mathbb{R}^{2,2}$ where, 
\begin{equation}
 B_\mu = e^{-\alpha} \epsilon_{\mu\nu\rho\sigma} X^\nu \partial_- 
X^\rho \partial_+ X^\sigma
\label{eq:AdS3_expr_for_B_Pohlmeyer}
\end{equation}   
with, 
\begin{eqnarray}
 \alpha & = & \ln(-\partial_+ X_\mu \partial_- X^\mu) 
\nn \eea 
We also define the quantitities, 
\bea
 u & = & B_\mu \partial_+^2 X^\mu \nn \\
 v & = & B_\mu \partial_-^2 X^\mu  \nn \eea
\paragraph{}
By virtue of this definition, as well as the equation of motion and
Virasoro constraint, the basis vectors are orthogonal and each vector
is either null or normalised
to unity. In particular we have,  
\begin{equation}
 B_\mu B^\mu = 1 \:, \qquad B_\mu X^\mu = B_\mu \partial_+ X^\mu = 
B_\mu \partial_- X^\mu = 0 \:,
\label{eq:def_vector_B_Pohlmeyer}
\end{equation}
\paragraph{}
The string equations of motion and Virasoro constraint can then be
transformed into equations for $\alpha$, $u$ and $v$. we find,  
\begin{equation}
 \partial_+\partial_-\alpha + e^\alpha + uv e^{-\alpha} = 0
\label{eq:general_alpha_eom}
\end{equation}
together with the conditions $\partial_{-}u=\partial_{+}v=0$ which
imply that we can set $u=u(\sigma^{+})$ and $v=v(\sigma^{-})$. Freedom
to choose these functions of the left and right moving coordinates
corresponds to the residual gauge invariance of the string left after
fixing conformal gauge.  
\paragraph{}
Now we introduce the following change of variable:
\begin{equation}
 \hat{\alpha} = \alpha - \frac{1}{2} \ln |u||v|
\label{eq:def_alphahat_Pohlmeyer}
\end{equation}
and define rescaled spacetime coordinates 
$\hat{\sigma}^{\pm}$ so that, 
 \begin{equation}
 \partial_+ = \sqrt{2|u|} \hat{\partial}_+ \:, \qquad \partial_- = 
\sqrt{2|v|} \hat{\partial}_-
\label{eq:transf_from_dpm_to_dpmhat}
\end{equation}
Finally the remaining equation (\ref{eq:general_alpha_eom}) takes the
form,  
\begin{equation}
 \hat{\partial}_+ \hat{\partial}_- \hat{\alpha} + 
\frac{1}{2} [e^{\hat{\alpha}} + \mathrm{sgn} (uv) e^{-\hat{\alpha}}] = 0
\label{eq:general_alphahat_eom}
\end{equation}
which reduces to the sinh-Gordon equation when $uv<0$ and to the
cosh-Gordon equation when $uv>0$. The strategy for finding string
solutions is to start from known solutions of the sinh-Gordon
equation, 
$\hat{\alpha}(\hat{\sigma}_{+}, \hat{\sigma}_{-})$ and reconstruct the 
string coordinates by solving the auxiliary linear differential
equations arising from the Lax formulation of the sinh-Gordon theory 
\cite{Jev1}.    
\paragraph{}
We now review the explicit solutions constructed in \cite{Jev1}. 
For convenience, from now on we will work in the worldsheet coordinates $\sigma$ and $\tau$, without
introducing the transformation \eqref{eq:transf_from_dpm_to_dpmhat} (all the solutions we are going to consider
have $u = 2$, $v = -2$ and thus, in these coordinates, the sinh-Gordon equation becomes
$\partial_+ \partial_- \hat{\alpha} + 4 \sinh \hat{\alpha} = 0$).
The simplest case corresponds to the vacuum solution of the
sinh-Gordon equation, 
\begin{equation}
 \hat{\alpha}_0 = 0 \quad \mathrm{or} \quad \alpha_0 = \ln 2
\label{eq:def_sinhG-vacuum}
\end{equation}
which yields the solution for the complex coordinates, 
\begin{eqnarray}
 Z_1 & = & X_0 + i X_1 = \cosh\rho \; e^{i t} \nonumber \\
 Z_2 & = & X_2 + i X_3 = \sinh\rho \; e^{i \phi} \nn
\end{eqnarray}
which takes the form, 
\begin{eqnarray}
 Z_1 & = & e^{i\tau} \cosh \sigma \nonumber \\
 Z_2 & = & e^{i\tau} \sinh \sigma
\label{vac}
\end{eqnarray}
This corresponds to the infinite spin limit of the GKP \cite{GKP} 
folded spinning
string in $AdS_{3}$. 
For coordinate range $\sigma \in
(-\infty,+\infty)$, it describes a straight line passing through the
centre of $AdS_3$, extending up to the boundary, which rigidly rotates 
at constant angular velocity. 
One can glue two of these lines
together in order to construct a folded closed string.
\paragraph{}
The energy and angular momentum of this solution are infinite. Hence
we will regulate the problem by considering instead a closed string of finite
length with folds at radial distance $\rho= \Lambda>>1$. Up to
subleading corrections, this corresponds to
the same solution but with the range of the worldsheet coordinate
restricted as $-\Lambda\leq \sigma \leq +\Lambda$. We then find, 
\begin{eqnarray}
 \Delta & = &  \frac{\sqrt{\lambda}}{\pi} 
\int_{-\Lambda}^{\Lambda} d\sigma \cosh^2 \sigma \,\,\, 
=\,\,\, \frac{\sqrt{\lambda}}{4\pi} 
\left( e^{2\Lambda} + 4 \Lambda - e^{-2\Lambda} \right) \nonumber \\
 S & = &  
\frac{\sqrt{\lambda}}{\pi}
\int_{-\Lambda}^{\Lambda} d\sigma \sinh^2 \sigma \,\,\, = \,\,\, 
\frac{\sqrt{\lambda}}{4\pi} \left(
    e^{2\Lambda} - 4 \Lambda - e^{-2\Lambda} \right) 
\nonumber \\
\label{ev}
\end{eqnarray}
Subtracting the regulated string energy and angular momentum we obtain
the standard formula, 
\begin{eqnarray}
E_{0}\,\,=\,\, \Delta - S & = & 
\frac{\sqrt{\lambda}}{\pi} 2\Lambda \nn \\ & = & 
\frac{\sqrt{\lambda}}{\pi}\left[ \log \left( \frac{\pi
     S}{\sqrt{\lambda}} \right) + 2\log 2 + 
O\left(\frac{\log S}{S}\right)\right] 
\label{eq:E-S_for_vacuum}
\end{eqnarray}
\paragraph{}
In this paper we are interested in classical excitations of the long
string solution which contribute a finite amount to the anomalous
dimension $E=\Delta-S$. 
As the long string corresponds to the vacuum solution of
the sinh-Gordon equation we expect that the configurations we seek
should correspond to excitations of the sinh-Gordon vacuum. 
One such excitation corresponds to the small fluctuations of the
sinh-Gordon field which carry energy of order one. Here we will focus
instead on states which have energy $E\sim \sqrt{\lambda}$ and
are visible as classical solutions of the string worldsheet theory. 
In fact, the
sinh-Gordon equation has singular soliton solutions and these are
natural candidates for the states we seek. The solution
describing a single 
(anti-)soliton moving at velocity $v$ is given by, 
\begin{equation}
 \alpha_{s/\bar{s}} = \ln 2 \pm \ln[\tanh^2 (\gamma (\sigma - v \tau))]
\label{eq:def_sinhG_1s1a}
\end{equation}
where the positive and negative signs correspond to the choice of
soliton ($s$) or anti-soliton ($\bar{s}$) respectively. Note that these 
configurations of the sinh-Gordon field $\hat{\alpha}$ are singular so it is
not obvious that they should be thought of as fluctuations around the
vacuum state. However, they do not necessarily lead to a singularity
in the string solution. In fact, the relation,
\begin{eqnarray}
-\partial_+ X_\mu \partial_- X^\mu & = & \exp(\alpha) 
\nn \eea 
shows that the singularity of the soliton solution
simply corresponds to a zero of the quantity 
$\partial_+ X_\mu \partial_- X^\mu$. Recalling the Virasoro
constraint (\ref{eq:Virasoropm}), we see that such a zero can occur at
a cusp point of the string where derivatives
$\partial_{\sigma}X_{\mu}$ vanish. The string solution is regular at
such a point and worldsheet densities of conserved
charges remain finite. As we see below, it seems that no such 
benign interpretation exists for the anti-soliton solution in the
present context.  
\paragraph{}
The sinh-Gordon equation is
exactly integrable and, as a consequence, one can find exact solutions
describing the scattering of any number of solitons and
anti-solitons. For example, solutions describing the scattering of 
two solitons or two anti-solitons are:
\begin{eqnarray}
 \alpha_{ss,\bar{s}\bar{s}} & = & 
\ln 2 \pm \ln \left[ \frac{v \cosh X - \cosh T}{v \cosh X + \cosh T} 
\right]^2 \nonumber \nn \eea
where $X=2\gamma\sigma$, $T=2 v \gamma \tau$ and $0<v<1$. The 
solution is given here in the centre of mass frame where the (anti-)solitons
have equal and opposite 
velocities $\pm v$. As in the more conventional case of the
sine-Gordon equation, the interaction between two solitons and between
two anti-solitons is repulsive while that between a soliton and an
anti-soliton is attractive and also gives rise to a classical boundstate or 
breather solution.  
\paragraph{}
Following \cite{Jev1}, one can construct string solutions
corresponding to an arbitrary number of solitons and anti-solitons
(and breathers) using the inverse scattering method. 
However we are primarily interested in
solutions which correspond to local excitations of the vacuum configuration 
(\ref{vac}). In
particular, we demand that the resulting string solutions asymptote to
the vacuum solution (\ref{vac}) as $\sigma\rightarrow \pm
\infty$ and also that the contribution of the excitation to the energy
$E=\Delta-S$ is finite. In fact the only solution presented in
\cite{Jev1} with the required property is the solution for the
scattering of two solitons with non-zero velocities $\pm v$, 
\begin{eqnarray}
 Z_1^{ss} & = & e^{i\tau} \frac{v \mathrm{ch} T \mathrm{ch} \sigma +
   \mathrm{ch} X \mathrm{ch} \sigma - \sqrt{1-v^2} 
\mathrm{sh} X \mathrm{sh}
  \sigma - i \sqrt{1-v^2} \mathrm{sh} T \mathrm{ch} \sigma}
{\mathrm{ch} T + v \mathrm{ch} X} \nonumber \\
 Z_2^{ss} & = & e^{i\tau} \frac{v \mathrm{ch} T \mathrm{sh} \sigma +
   \mathrm{ch} X \mathrm{sh} 
\sigma - \sqrt{1-v^2} \mathrm{sh} X \mathrm{ch}
  \sigma - i \sqrt{1-v^2} \mathrm{sh} T \mathrm{sh}
  \sigma}{\mathrm{ch} T + v \mathrm{ch} X} 
\nonumber \\
\label{eq:2s_solution}
\end{eqnarray}
where $\mathrm{ch} X \equiv \cosh X$ and $\mathrm{sh} X \equiv \sinh
X$ and $0<v<1$. Note that the global time $t$ and the worldsheet time
$\tau$ are different for the above solution. 
Several plots of this solution at constant global time $t$ are shown in
Fig. \ref{fig:Plot_2s_for_6_values_of_t}. This solution has two small
spikes located at the positions of the two solitons, which start at
the endpoints of the string, then approach each other until they
scatter at the origin and then move away towards the
endpoints. Interestingly the same solution plotted at constant worldsheet
time $\tau$ has no cusps.
\begin{figure}%
\includegraphics[width=0.83\columnwidth]{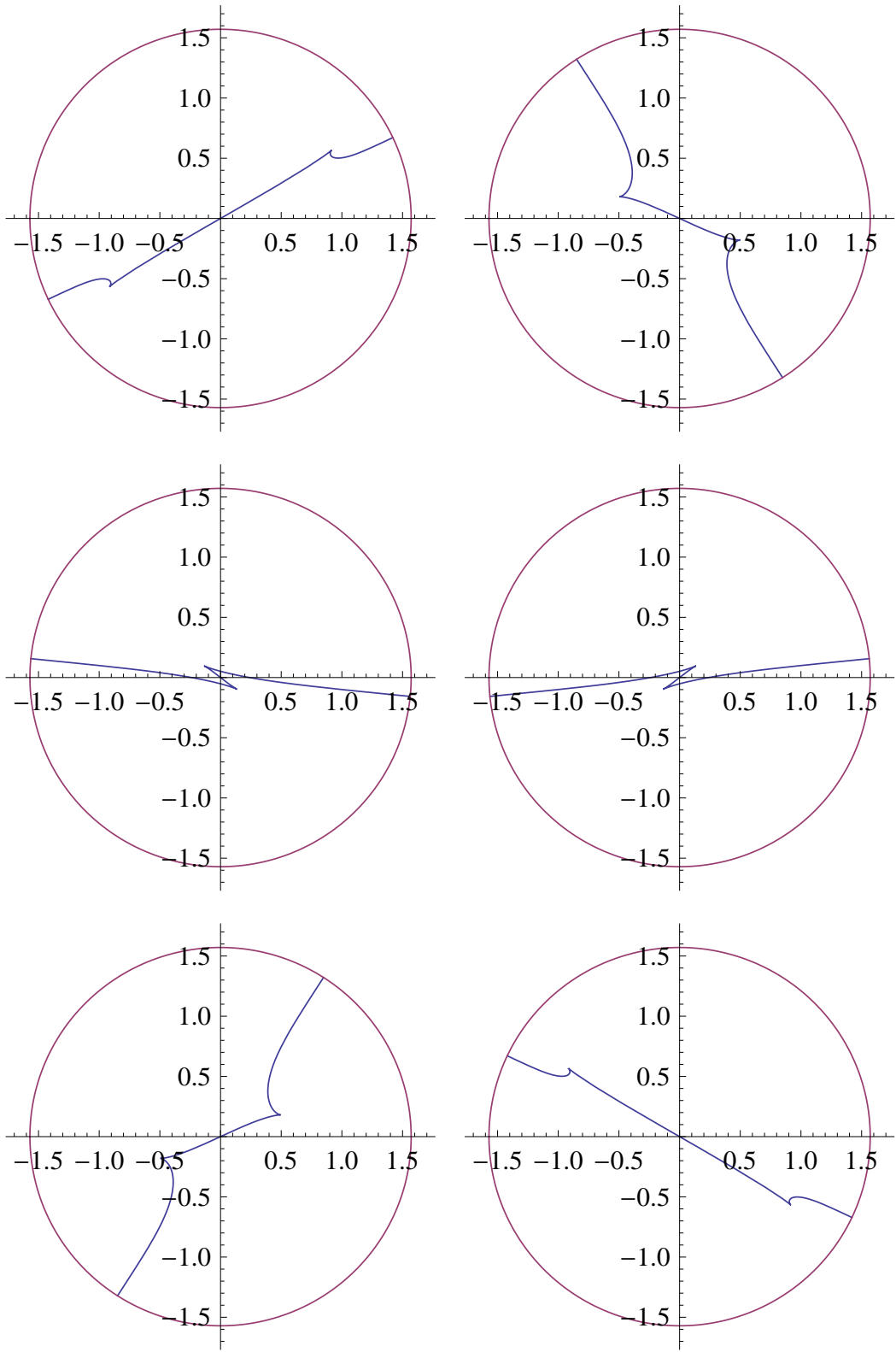}%
\caption{The 2-soliton solution at $t=-2.7,-1,-0.1,0.1,1,2.7$ (from
  left to right and top to bottom). At $t=0$ the string coincides 
with the horizontal axis.}%
\label{fig:Plot_2s_for_6_values_of_t}%
\end{figure}
The solution (\ref{eq:2s_solution}) has asymptotics, 
\bea 
Z_{1} & \simeq & \frac{1}{2}e^{\pm \sigma}\,e^{i\tau}\left(
  \frac{1-\sqrt{1-v^{2}}}{v}
\right)  \nn \\ 
Z_{2} & \simeq & \pm \frac{1}{2}e^{\pm \sigma}\,e^{i\tau}\left(
  \frac{1-\sqrt{1-v^{2}}}{v}
\right)  \label{asy} \eea
as $\sigma\rightarrow\pm \infty$ which indeed match those of the
vacuum configuration (\ref{vac}) up to a finite shift in the radial
coordinate $\rho$ which will be important in the following. 
\paragraph{}
For the string solution corresponding to two anti-solitons, the
denominators in (\ref{eq:2s_solution}) are replaced by 
$\mathrm{ch} T - v \mathrm{ch} X$ while in the solution for
soliton-antisoliton scattering the relevant denominator
is\footnote{See equations (4.41) and (4.42) of \cite{Jev1}} 
$\mathrm{sh} T \pm v \mathrm{sh} X$. The denominator for the breather
solution then follows from analytic continuation to imaginary velocity
$v=iw$. All of these denominators have zeros at finite values of
$\sigma$ leading to poles in the corresponding solution which yield an
infinite contribution to the energy. Hence, configurations involving
anti-solitons do not seem to correspond to local excitations of the 
vacuum (\ref{vac}).
\paragraph{}
As for the vacuum solution, we 
will consider the solution (\ref{eq:2s_solution}) 
as one half of a long folded string extending to radial
coordinate $\rho=\Lambda>>1$. The asymptotics (\ref{asy}) then imply we
must restrict the spatial
worldsheet coordinate to the range $-\tilde{\Lambda}\leq \sigma \leq 
+\tilde{\Lambda}$ with $\tilde{\Lambda}=\Lambda-\delta\Lambda$ where, 
\bea 
\delta\Lambda & = & \log\left(
  \frac{1-\sqrt{1-v^{2}}}{v}\right) 
\,\,\,=\,\,\, \frac{1}{2}\log\left(\frac{1-\sqrt{1-v^{2}}}
{1+\sqrt{1-v^{2}}}\right) \nn \eea
In the second equality we have chosen the branch of the logarithm such
that the expression is even under $v \rightarrow -v$. This is
appropriate because the solution is invariant under the interchange of
the two identical solitons.         
The shift $\delta\Lambda$ 
reflects the contribution of the excitations to the total
length of the string. Note that it vanishes for $v=1$ where the two
soliton solution simply reduces to the vacuum solution (\ref{vac}). On
the other hand, the additional contribution diverges as 
$v\rightarrow 0$ indicating a change of the asymptotic behaviour of
the string solution in this limit\footnote{In this context, 
it is worth noting that the end
points of the folded string corresponding to the regulated vacuum
configuration discussed above themselves correspond to sinh-Gordon solitons 
with zero velocity \cite{Jev1}. 
As a soliton velocity goes to zero we expect that
it corresponds to a new asymptotic region of the string which
approaches the boundary, leading to a string solution with an 
additional spike of the type discussed in \cite{Kruc1}.}.   
\paragraph{}
The resulting string energy $E=\Delta -S$ can be computed directly
from the regulated solution. 
Relative to the vacuum solution, we replace one
half of the closed 
string by the corresponding two soliton solution to get, 
\begin{eqnarray}
 E & = & \frac{1}{2}E_{0} + \frac{\sqrt{\lambda}}{2\pi} 
\int_{-\tilde{\Lambda}}^{\tilde{\Lambda}} 
d\sigma \frac{1 - 3 v^2 + \cosh 2T + v^2 \cosh 2X}{2(\cosh T + v \cosh X)^2}
  \nonumber \\
 & = & \frac{1}{2}E_{0}+
\frac{\sqrt{\lambda}}{2\pi} \left[ \sigma - 
\frac{v \sqrt{1-v^2} \sinh X}{\cosh T + v \cosh X}
\right]_{-\tilde{\Lambda}}^{\tilde{\Lambda}} 
\nonumber \\
 & = & \frac{\sqrt{\lambda}}{2\pi} \left[ 2\Lambda+2\tilde{\Lambda} - 
2 \sqrt{1-v^2} + O(e^{-2\gamma\Lambda}) \right] \nn \\ & = &
E_{0}\,\, + \,\, 2E_{\rm sol}(v) \,\, + \,\,O(e^{-2\gamma\Lambda}) \nn
\eea 
where, 
\bea 
E_{\rm sol}(v) & = &
\frac{\sqrt{\lambda}}{2\pi}\left[-\delta\Lambda\,\,-\,\,\sqrt{1-v^{2}}
\right] \nn \\ & = & 
\frac{\sqrt{\lambda}}{2\pi}
\left[\frac{1}{2}\log\left(\frac{1+\sqrt{1-v^{2}}}
{1-\sqrt{1-v^{2}}}\right)\,\,-\,\,
\sqrt{1-v^{2}}\right] \label{de} \eea 
is naturally interpreted as the energy of a single soliton of velocity
$-1\leq v \leq +1$. Note that $E_{\rm sol}$ diverges as the soliton
velocity $v$ goes to zero reflecting the divergent contribution to the
length of the string mentioned above.   
\paragraph{}
As the consitituent solitons of the two soliton scattering solution 
are well seperated at very early and late times, it is intuitively
clear that string solutions corresponding to individual solitons with
vacuum asymptotics must also exist. 
Indeed such solutions were also presented in \cite{Jev1}
(see equations (4.25, 4.26)), but were found to have infinite energy. 
They also have different asymptotics to the vacuum configuration
studied above\footnote{In particular, the asymptotic values of the angular
coordinate $\phi$ are the same at both ends of the string while they
differ by $\pi$ in the vacuum solution}. In fact this pathological
behaviour arises because the solution (4.25, 4.26)
of \cite{Jev1} corresponds to a single soliton with zero velocity and
is related to the divergence of $E_{\rm sol}(v)$ as $v\rightarrow 0$
found above. A solution corresponding to a single soliton with 
non-zero velocity
$v>0$ can be obtained by space and time translation of the two-soliton 
scattering 
solution so that one cusp is located near the origin and the other is sent to
infinity. We construct such a solution in the Appendix 
and show that the solution
indeed has vacuum asymptotics and energy $E_{0}+E_{\rm sol}$ as
expected for a single soliton. 
\paragraph{}
Having established the existence of excitations of finite energy we
now want to determine their dispersion relation. In particular, as the
cusps move along the string with velocity $v$ as measured in the
spacelike worldsheet coordinate $\sigma$, we want to identify the
conserved momentum $P_{\rm sol}(v)$ which is canonically conjugate to the 
position of the soliton in these coordinates. Here we will follow
the same line of reasoning used for the case of Giant Magnons in
\footnote{In particular, see discussion around eqns (2.16-2.19) of this 
reference.} \cite{HM}. Consider a configuration with $M$ cusps located
at positions $\sigma=\sigma_{l}(\tau)$ in the worldsheet coordinate $\sigma$
introduced above, moving with velocities $v_{l}=d\sigma_{l}/d\tau$
for $l=1,\ldots, M$. The total energy of the configuration is,   
\bea
E=E_{0}(S)\,\,+\,\,\sum_{l=1}^{M}\,E_{\rm sol}(v_{l}) 
\label{total} \eea  
The energy $E=\Delta-S$ is canonically conjugate to the global coordinate
$\tilde{t}=(t+\phi)/2$ and we can define a canonical momentum $P_{l}$
for each soliton via Hamilton's equation,   
\begin{eqnarray}
\frac{d{\sigma}_{l}}{d\tilde{t}} & = &  \,\,
\frac{\partial E}{\partial P_{l}} \label{Ham}
\eea
\paragraph{}
An important subtlety is that the global time $\tilde{t}$ appearing in
the above equation is not equal to the worldsheet time in the string
solutions considered above. However they are equal in the vacuum
solution (\ref{vac}) and, as each 
sinh-Gordon soliton is localised, we have  
$d\tilde{t}/d\tau\rightarrow 1$ exponentially fast away from the centre 
of each cusp. Thus the differential 
$\tilde{v}_{l}=d{\sigma}_{l}/d\tilde{t}$ appearing in (\ref{Ham}) will be
equal to the worldsheet velocity $v_{l}$ 
up to exponentially small corrections for
almost all times\footnote{This only fails to be true during a finite time
  interval of duration of order $1/v$ when the soliton crosses the
  origin. This effect will produce a subleading correction to the
  semiclassical spectrum discussed below.}. 
Making the replacement $\tilde{v}\rightarrow v$ in the  
Hamilton equation (\ref{Ham}) 
for a single soliton moving at constant velocity $v$ we get,  
\bea 
v\,\, = \,\, 
\frac{dE_{\rm sol}}{dP_{\rm sol}} & = & \frac{dE_{\rm
      sol}}{dv}/
\frac{dP_{\rm sol}}{dv} \nn \eea 
or equivalently, 
\bea
\frac{dP_{\rm sol}}{dv} \,\,=\,\,\frac{1}{v}\,\frac{dE_{\rm sol}}{dv} & = & 
- \frac{\sqrt{\lambda}}{2 \pi} \frac{\sqrt{1-v^2}}{v^{2}} \label{int2} \eea 
where we used (\ref{de}) in the second equality.
As the soliton solutions considered above revert to the vacuum for
$|v|=1$ we integrate (\ref{int2}) with boundary condition 
$P_{\rm sol}(\pm 1)=0$ to get,  
\bea 
P_{\rm sol}(v) & = &
\frac{\sqrt{\lambda}}{2\pi}\left[\frac{\sqrt{1-v^{2}}}{v}\,\,-\,\,
{\rm Tan}^{-1}\left(\frac{\sqrt{1-v^{2}}}{v}\right)\right] \label{dp} \eea
Equations (\ref{de}) and (\ref{dp}) 
constitute the dispersion relation of the soliton. 
Notice that the conserved momentum $P_{\rm sol}(v)$ is an odd function
of the velocity $v$ by construction. Thus the total momentum of the
two soliton solution considered above is zero. More generally we
might expect that an $M$-soliton closed string solution should obey a 
level-matching condition of the form, 
\bea 
P &= & \sum_{l=1}^{M}\, P_{\rm sol}(v_{l})\,\,=\,\,0 \nn \eea 
As mentioned above, the 
situation is complicated by the fact that the folds at the
end of the string themselves correspond to solitons with zero velocity which
therefore yield two infinite contributions to the total momentum of
opposite sign which can cancel up to a finite remainder. 
This consideration presumably allows for the existence of the
one-soliton excitation of the folded string discussed in the Appendix.  
\paragraph{}
As mentioned above, the physical content of the dispersion
relation is contained in the semiclassical spectrum of excitations
around the vacuum configuration.
Classically the spin $S$ and soliton velocities $v_{l}$ are continuous
parameters of the solution. In leading order semi-classical
quantization $S$ is naturally quantized in integer units. In addition,
the semiclassical wavefunction for a soliton of velocity $v$ takes the
form, 
\bea 
\Psi(\sigma) & = & \exp\left(iP_{\rm sol}(v)\sigma\right)
\nn \eea
The quantization condition for the soliton velocity comes from
demanding that the wavefunction should be invariant under the shift
$\sigma\rightarrow \sigma+ 2\Lambda+2\tilde{\Lambda}$
corresponding to a closed string of total length 
$L=2\Lambda + 2\tilde{\Lambda}$. For our purpose the leading order
behaviour $L\simeq 2\log(S)+{\rm O}(S^{0})$ is
sufficient to find the leading-order quantization
condition\footnote{In general, there are additional corrections
  coming from the two-body scattering of solitons leading to a
  quantisation condition of Bethe
  Ansatz type. However, we will restrict our attention to the case
  where the quantized momentum 
$P_{\rm sol}$ remains of order one as $S\rightarrow \infty$  
and the scattering phase is subleading. A similar correction from the
inequality of global and worldsheet time near each soliton arises at the
same order as the scattering phase.}   
\bea
P_{\rm sol}(v)\cdot 2\log S & \in & 2\pi\,\,\mathbb{Z} \label{quant} \eea    
Implementing these quantization conditions for each soliton
contribution to the total
energy (\ref{total}) yields a definite prediction for the
semiclassical spectrum. In a forthcoming paper we will reproduce this 
spectrum directly from the finite-gap formalism of \cite{KMMZ}.      
\paragraph{}
It is interesting to compare our string theory results with the
corresponding spectrum of excitations in the dual gauge theory. 
The dual to the GKP spinning string lies in the $SL(2)$ sector of
the ${\cal N}=4$ theory consisting of 
operators of the form, 
\bea 
\hat{O} & \sim & {\rm Tr}_{N}\left[\,\mathcal{D}_{+}^{s_{1}}Z\,
\mathcal{D}_{+}^{s_{2}}Z\,\ldots\, \mathcal{D}_{+}^{s_{J}}Z\,\right] 
\label{sl2} 
\eea
which have Lorentz spin  $S = \sum_{l=1}^{J}\,\,s_{l}$ and twist $J$.  
Here $\mathcal{D}_{+}$ is a light-cone covariant derivative in
Minkowski space. The one-loop anomalous dimensions of these operators are
determined by the energy levels of an integrable ${\rm SL}(2,\mathbb{R})$ 
spin chain which are determined exactly by the Bethe ansatz. To
formulate the Bethe ansatz one starts from the ferromagnetic vacuum of
the spin chain which corresponds to the chiral primary $\hat{O}_{\rm
  BPS}={\rm Tr}_{N} Z^{J}$. 
Excitations around this groundstate corresponding to insertion of
impurities $\mathcal{D}_{+}$, which carry one unit of spin 
$S$, and are known as magnons. These excitations
each carry a conserved momentum and also 
obey an exclusion principle (see eg \cite{Faddeev}) 
which forbids any two magnons occupying
the same state. The state of fixed spin with lowest energy 
$E=\Delta-S$ is obtained by filling the Dirac sea with magnons. 
However, for an operator of twist $J$ there are always exactly 
$J$ holes in the distribution of mode numbers \cite{BGK,FS}.  
\paragraph{}
In the $S\rightarrow \infty$ limit, the number of magnons becomes
large but the number of holes $J$ remains fixed. Remarkably,
the holes acquire the attributes of particles. 
Specifically, each hole carries conserved energy $E_{\rm hole}(u)$
parametrised in terms of a complex rapidity $u$ as \cite{BGK},     
\bea E_{\rm hole}(u)= \frac{\lambda}{8\pi^{2}}\left[\psi(1/2
    +iu) +\psi(1/2- iu)-2\psi(1) \right] \label{disp12} \eea 
where $\psi(x)=d(\log\Gamma(x))/dx$. Further, the total energy of the
state is essentially the sum of the energies of the individual
holes, 
\bea E & = & \frac{\lambda}{4\pi^{2}}\, \log 2 \,\,+\,\,\
\sum_{j=1}^{J}\, E_{\rm hole}(u_{j}) \nn
\eea
In the groundstate configuration which is dual to the spinning
string there are two "large" holes with rapidities 
$u_{1}$, $u_{2}$ $\sim S$, whose contribution
dominates the energy leading to the standard result, 
\bea
E_{0}=\frac{\lambda}{2\pi^{2}}\, \log S \,\,\,+\,\,\,
{\rm O}\left(S^{0}\right) \label{e0} 
\eea
where we have used the asympototic form $\psi(x)\sim \log x$ for large
$x$. This gauge theory formula famously agrees with the string theory
formula \cite{GKP} (\ref{eq:E-S_for_vacuum}) 
up to the replacement of the prefactor in
both formulae by $2\Gamma(\lambda)$ where $\Gamma$ is the cusp
anomalous dimension.  
\paragraph{}
The rapidities of the remaining $J-2$ holes are quantised according to
a dual Bethe ansatz equation (see equation (3.41) in \cite{BGK}). In
the large-$S$ limit, the quantization condition for the hole 
rapidities takes the form, 
\bea 
8u_{j}\cdot \log S & \in & 2 \pi\,\mathbb{Z} \nn \eea
for $j=3\ldots J$. Comparing with the quantization condition
(\ref{quant}) we identify the hole momentum according to 
$P_{\rm hole}(u)=4u$. 
\paragraph{}
In the groundstate, the lowest allowed values of $u$ 
are occupied which gives $u_{j}\sim 1/\log S$ for $j=3,\ldots
J$. An excitatation of the groundstate is
obtained by allowing one of the holes to have rapidity $u_{j}$ of
order one. The resulting state can be regarded as a localised excitation
of the groundstate with energy $E_{\rm hole}(u_{j})$ given by (\ref{disp12})
and momentum $P_{\rm hole}(u_{j})=4u_{j}$. In general the resulting
dispersion relation is quite different from the string theory
result. However, for large momentum $u_{j}>>1$ we obtain the leading
order result, 
\bea 
E_{\rm hole} & = & \frac{\lambda}{4\pi^{2}}\,\log\,|P_{\rm
  hole}|\,\, + \,\,  {\rm
  O}\left(P^{0}_{\rm hole}\right)    
\nn \eea   
With the standard replacement of $\lambda/4\pi^{2}$ by the cusp
anomalous dimension $\Gamma(\lambda)$, this 
precisely matches the large momentum form (\ref{largep}) of the 
string theory dispersion relation (\ref{disp2}). It is natural
therefore to conjecture that the solitonic excitations we consider are
dual to the gauge theory holes. This is also consisitent with the map
between the classical gauge theory spin chain and spiky strings
proposed in \cite{DS,DL1}.   
\paragraph{}
As mentioned above, the dispersion relation for the gauge theory
solitons becomes relativistic at low energy and momenta. In particular 
it corresponds to the dispersion relation of a massless relativistic 
particle on the string worldsheet or, more precisely, a particle whose mass is 
small compared to $\sqrt{\lambda}$ so that the energy gap is zero
in the leading semiclassical approximation. 
At low energy
and momentum the string also has the usual spectrum of quadratic
fluctuations. A string moving in $AdS_{3}$ has only one transverse
mode. This mode has a non-zero mass $m^{2}=4$ \cite{AM2} which is 
determined by the spontaneous breaking of the  
$SL(2,\mathbb{R})$ "symmetry" 
\footnote{This is a novel version of the Goldstone
  phenomenon where the spontaneous breaking of a charge which has
  non-zero commutator with the Hamiltonian leads to a {\em massive} boson 
\cite{AM2}}. 
The appearance of a second mode with small or zero energy gap is
potentially puzzling so it is natural to conjecture that the two
excitations are continuously related to each other in the same sense
that Giant Magnons are related to the quadratic fluctuations of the
BMN string groundstate. If this is the case, quantum corrections must
modify the dispersion relation (\ref{disp2}) to introduce the
appropriate mass term at low energy. An important difference to the
Giant Magnon case is that, here, supersymmetry is strongly broken by the 
groundstate. Correspondingly the dispersion relation 
considered here is not constrained by supersymmetry or, at least,
there is no obvious candidate for an exact dispersion relation. 
\paragraph{}
There are many interesting questions for further study. It is
straightforward to calculate the semiclassical S-matrix for the Giant
Holes studied in this paper and to formulate a leading order Bethe
ansatz for the system.  It also would be
interesting to investigate the dispersion relation of holes at
intermediate values of the coupling using the known Asymptotic Bethe Ansatz 
and verify directly the 
interpolation to the worldsheet solitons studied here. The results and
techniques of \cite{BFR} should be useful in this context. Finally one
could also consider how the Giant Holes are embedded in the full
$AdS_{5}\times S^{5}$ background. The full spectrum of small
fluctuations around the spinning string was identified in \cite{AM2}. 
It seems possible that each of these modes is continuously related to a
solitonic solution. Once embedded in the full 
background we expect to find additional states associated with the
zero modes of solitons. As in the case of Giant Magnons, this should
give rise to different "polarisation states" of the Giant Hole in
one-to-one correspondence with the quantum numbers of the elementary
fluctuations around the long spinning string. We hope to investigate
these questions in the near future.      



\paragraph{}

The authors would like to thank Juan Maldacena 
for helpful comments.

\section*{Appendix: One-soliton solution}
\paragraph{}
A solution corresponding to a single soliton with non-zero velocity
$v>0$ can be obtained by space and time translation of the two-soliton 
scattering
solution so that one cusp is located near the origin and the other is sent to
infinity. Applying this proceedure to (\ref{eq:2s_solution}), we find the
following solution of the string equation of motion and Virasoro
constraints,      
\begin{eqnarray}
 Z_1 & = & e^{i\tau} \frac{e^\Sigma (\cosh\sigma - \sqrt{1-v^2} \sinh
   \sigma) + e^{-\Sigma} (v - i \sqrt{1-v^2}) 
\cosh \sigma}{v
  e^\Sigma + e^{-\Sigma}} \nonumber \\
 Z_2 & = & e^{i\tau} \frac{e^\Sigma (\sinh\sigma - \sqrt{1-v^2} \cosh
   \sigma) + e^{-\Sigma} (v - i \sqrt{1-v^2}) 
\sinh \sigma}{v
  e^\Sigma + e^{-\Sigma}}
\label{eq:1s_solution_Rmover_tau+inf}
\end{eqnarray}
where $\Sigma = (X-T)/2 = \gamma (\sigma - v\tau)$, $X = 2\gamma
\sigma$, $T =  2v\gamma\tau$ and $\gamma = \gamma (v) = (1-v^2)^{-
  \frac{1}{2}}$. One may easily check the corresponding sinh-Gordon
angle $\alpha$ coincides with that of the one-soliton solution
(\ref{eq:def_sinhG_1s1a}) with velocity $v$, up to a constant shift in $\tau$.
\paragraph{}
The solution (\ref{eq:1s_solution_Rmover_tau+inf}) has asymptotics, 
\bea 
Z_{1} & \simeq & \frac{1}{2}e^{\sigma}\,e^{i\tau}\left(
  \frac{1-\sqrt{1-v^{2}}}{v}
\right)  \nn \\ 
Z_{2} & \simeq & \frac{1}{2}e^{\sigma}\,e^{i\tau}\left(
  \frac{1-\sqrt{1-v^{2}}}{v}
\right)  \label{asy2} \eea
as $\sigma\rightarrow\infty$ and, 
\bea 
Z_{1} & \simeq & \frac{1}{2}e^{-\sigma}\,e^{i\tau}
\left(v-i\sqrt{1-v^{2}}
\right) \nn \\ 
Z_{2} & \simeq & -\frac{1}{2}e^{-\sigma}\,e^{i\tau}
 \left(v-i\sqrt{1-v^{2}}
\right)  \label{asy3} \eea
as $\sigma\rightarrow -\infty$. 
Or in terms of the global coordinates, 
\begin{eqnarray}
 t(\tau,\sigma) & \to & \begin{cases}
                         \tau \:, \quad \textrm{for } \sigma \to + \infty \\
                         \tau - \beta \:, \quad 
\textrm{for } \sigma \to - \infty
                        \end{cases} \nonumber \\
 \phi(\tau,\sigma) & \to & \begin{cases}
                            \tau \:, \quad 
\textrm{for } \sigma \to + \infty \\
                            \tau - \beta + \pi \:, \quad 
\textrm{for } \sigma \to - \infty
                           \end{cases} \nonumber \\ 
 \rho(\tau,\sigma) & \to & \begin{cases}
                         \sigma + \alpha 
\:, \quad \textrm{for } \sigma \to + \infty \\
                         - \sigma \:, \quad 
\textrm{for } \sigma \to - \infty
                        \end{cases}
\label{eq:behaviour_of_t_and_phi_as_sigma_to_pm_infty_1s_solution}
\end{eqnarray}
where $\beta = {\rm Tan}^{-1}(\sqrt{1-v^2}/v)$ and 
$\alpha=\log[(1-\sqrt{1-v^2})/v]$. 
\paragraph{}
The asymptotic behaviour of the string coordinates imply that, 
\begin{equation}
 \phi \to \begin{cases}
           t \:, \quad \textrm{for } \sigma \to + \infty \\
           t + \pi \:, \quad \textrm{for } \sigma \to - \infty
          \end{cases}
\label{eq:phi_at_endpoints_at_constant_t_1s_solution}
\end{equation}
from which we obtain $\Delta\phi = \lim_{\sigma \to +\infty} 
\phi(t,\sigma) - \lim_{\sigma \to -\infty} \phi(t,\sigma) = -\pi$, $\forall t$.
Thus the asymptotics match those of the vacuum solution. To
evaluate the energy we consider the solution as one half of a closed
folded string stretching to radial coordinate $\rho=\Lambda>>1$. Thus
we must restrict the range of the worldsheet coordinate $\sigma$
according to $-\Lambda \leq \sigma \leq \tilde{\Lambda}$. Where 
$\tilde{\Lambda}=\Lambda-\alpha$. The resulting string energy is, 
\begin{eqnarray}
 \Delta - S & = & \frac{1}{2}E_{0}\,\,+\,\,
\frac{\sqrt{\lambda}}{2\pi} \int_{-\Lambda}^{\tilde{\Lambda}} 
d\sigma \left[ 1 - 2 v e^T \frac{e^X}{(e^T + v e^X)^2} \right] \nonumber \\
       & = & \frac{1}{2}E_{0}\,\,+\,\,
\frac{\sqrt{\lambda}}{2\pi} \left[ \sigma +
         \frac{e^T}{\gamma(e^T + v e^X)} 
\right]_{-\Lambda}^{\tilde{\Lambda}} \nonumber \\
       & = & \frac{\sqrt{\lambda}}{2\pi} \left[ 3 \Lambda 
+ \tilde{\Lambda} - 
\sqrt{1-v^2} + O(e^{-2\gamma\Lambda}) \right]  \nn \\ 
& = &  E_{0}+E_{\rm sol}(v)\,\,+\,\, O(e^{-2\gamma\Lambda})
\label{eq:evaluating_E-S_analytically_for_1s_solution}
\end{eqnarray}
as expected.

\end{document}